# Enhancing Microwave Heating Uniformity in Cavities Using a 2-Bit Coding Metasurface

Zhongyin Peng, *Graduate student member, IEEE*, Chengrong Wang, Changjun Liu, *senior Member, IEEE*, Xiang Zhao, and Liping Yan, *senior Member, IEEE*

*Abstract*—A novel method for enhancing microwave heating uniformity using a 2-bit coding metasurface is proposed. This metasurface is specially designed to scatter incident waves into multiple directions at 2.45 GHz rather than just one, significantly improving the electric field distribution uniformity within a cavity, eliminating the need to redesign the cavity itself or modify the power excitation. Simulated results demonstrate a reduction in the coefficient of variation (COV) of potato temperature from 0.694 to 0.461, indicating enhanced heating uniformity while maintaining efficiency. Additionally, the measured potato temperatures show excellent agreements with the simulated results, validating the effectiveness of the proposed method.

## I. INTRODUCTION

MICROWAVE heating has been widely applied in food processing, drying, chemical reactions, and medicine due to its high heating efficiency, low energy consumption, and environmental friendliness [1] - [3]. However, the non-uniform distribution of electric field (E-field) within microwave heating cavities presents a significant technical challenge [4]. This uneven distribution leads to issues such as hotspots and thermal runaway, which are major obstacles to large-scale industrial applications [5], [6].

Turntables [7] and mode stirrers [8] are two traditional methods for improving heating uniformity, which remain effective even today. In recent years, numerous methods have been proposed to further enhance heating uniformity. These methods can be categorized into two main types: boundary variation-based methods, such as optimizing irregular cavity walls [9] and loading rotary metallic columns array in the cavity [4], and microwave power excitation-changing methods, such as rotatable dual-port feeding [10], continuously adjustable multiple excitation sources [11], phase-shifting dual-port feeders [12], frequency stirring [13], and reconfigurable Fresnel zone plates for beam steering [14]. While these methods can effectively improve heating homogeneity, they may increase costs and workload or require redesigning the heating system.

An effective alternative to enhance microwave heating uniformity in a cavity is covering a metasurface on its wall [15], [16]. Metasurfaces are artificial subwavelength 2-D planar structures that can manipulate and control electromagnetic (EM) waves [17] - [23]. Our previous work demonstrated that a randomly arranged 1-bit coding metasurface can enhance the uniformity of the field inside the cavity [15].

In this letter, a 2-bit coding metasurface operating at 2.45 GHz is designed to improve heating uniformity in a microwave cavity. By designing a unit cell with a period of only 0.08 $\lambda_0$ (the wavelength of the center frequency) and optimizing the elements arrangement via a genetic algorithm (GA), the EM waves incident on the metasurface can be scattered into multiple directions rather than just one. The uniformity of the field distribution in the cavity is significantly improved, thereby enhancing the heating uniformity. Good agreements between the measurements and simulations validate this improvement. This novel metasurface-based method requires neither redesign of the cavity nor modification of power excitation and can be employed in any rectangular heating cavity operating at 2.45 GHz.

## II. METASURFACE-LOADED HEATING CAVITY

The heating uniformity in a microwave cavity is significantly influenced by the distribution of EM field, as the temperature distribution of a heated material is governed by the heat transfer equation in the solid medium.

$$\rho C_p \frac{\partial T}{\partial t} - k \nabla^2 T = P \tag{1}$$

The explanations of variables are given in [24], [25]. The absorbed microwave energy per unit volume is

$$P = \frac{1}{2} \omega \varepsilon_0 \varepsilon_r^{"} | \boldsymbol{E} |^2 \tag{2}$$

where $\omega$ is the angular frequency of the EM waves, $\varepsilon_0$ and $\varepsilon_r^{"}$ are the vacuum dielectric constant and the imaginary part of the complex relative permittivity, respectively, and $\boldsymbol{E}$ denotes the E-field intensity within the material. Consequently, a uniform field distribution results in homogenous heating. However, the intrinsic standing wave characteristics of a microwave cavity lead to a nonuniform field distribution. To tackle this without redesigning the cavity or altering the excitation, an effective yet simple approach is to modify the

This work was supported in part by the National Natural Science Foundation of China (NSFC) under Grant U22A2015 and in part by the Sichuan Science and Technology Program under Grant 2024YFHZ0282. *(Corresponding author: Liping Yan).*

Zhongyin Peng, Changjun Liu, Xiang Zhao, Liping Yan are with the College of Electronics and Information Engineering, Sichuan University, Chengdu, 610065, China. (e-mail: liping_yan@scu.edu.cn)

Chengrong Wang is with the College of Electronics and Information Engineering, Sichuan University, Chengdu, 610065 China, and Department of Electronic Engineering, Chengdu Jincheng College, Chengdu, 611731 China.

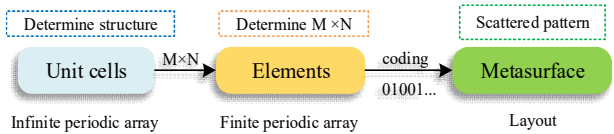

**Fig. 1.** Design diagram of a coding metasurface.

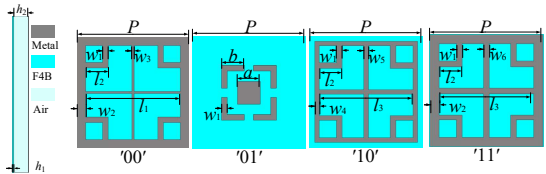

**Fig. 2.** The proposed 2-bit coding metasurface. $P = 10$, $h_1 = 1$, $h_2 = 16$; $l_1 = 8.4$, $l_2 = 1.5$, $l_3 = 8.2$, $w_1 = 0.5$, $w_2 = 0.8$, $w_3 = 0.2$, $w_4 = 0.4$, $w_5 = 0.4$, $w_6 = 0.5$, $a = 2$, $b = 2$. (unit: mm).

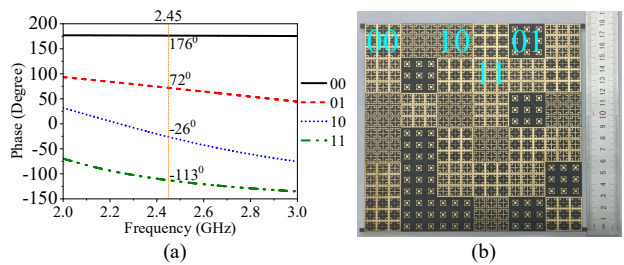

**Fig. 3.** The phase response corresponding to four unit cells (a) and the optimized arrangement of elements in the proposed metasurface (b).

cavity boundary by placing a metasurface on its wall. EM waves incident on metasurfaces satisfy the generalized Snell's law [26], enabling the metasurface to scatter the incident waves into multiple directions rather than just one, thereby resulting in enhanced uniformity of the field distribution.

*A. Elements Design and Arrangement*

Considering that 2-bit coding offers greater flexibility in manipulating EM waves compared to 1-bit coding [27], a 2-bit coding metasurface is proposed to modify the boundary in a microwave heating cavity. This 2-bit coding metasurface composes of $2^2$ elements: ′00′, ′01′, ′10′, and ′11′, which correspond to relative phase responses of 0, $\pi/2$, $\pi$, and $3\pi/2$, respectively, thereby ensuring the four states cover a full $2\pi$ phase range. By arranging these elements in specific sequences, the scattering pattern of EM waves is flexibly controlled. The design diagram is summarized in Fig. 1.

Current designs of coding metasurfaces for EM scattering are primarily used for radar cross section (RCS) reduction. Metallic patch and square ring have been frequently used as the unit cells structure with a period of about 0.2 $\lambda_0$ and even 0.5 $\lambda_0$ [27], [28]. However, these elaborated designs cannot be directly applied for microwave heating in a 2.45 GHz cavity due to constraints posed by the cavity's dimensions and the unit cell size. In this work, a 2-bit coding metasurface design is proposed for the specific cavity dimensions (291 mm × 291 mm × 186 mm) with a metasurface size of 180 mm × 180 mm.

Considering that flexible control of EM waves demands enough elements [28], the metasurface is composed of 6 × 6

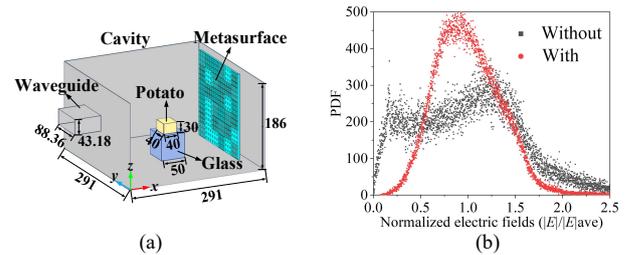

**Fig. 4.** (a) Structure and dimensions of a metasurface-loaded cavity (unit: mm) and (b) PDFs of normalized E-field distributions in the cavity without and with the metasurface at back of wall.

elements with each containing 3 × 3 unit cells to mitigate phase distortions. The period of unit cells is, thus, set as 10 mm, corresponding to 0.08 $\lambda_0$. The unit cells of square patch and ring with limited structural parameters, cannot provide the required phase responses for 2-bit coding metasurface with this small period. Therefore, a novel structure is introduced, incorporating a square ring with a cross and four L-shaped branches in the metallic patterned layer, as illustrated in Fig. 2. A ground plane is separated from the patterned layer by a 16 mm air gap. The metallic patterns are printed on a 1 mm-thick F4B dielectric substrate with a dielectric constant of 2.65 and a loss tangent of 0.001. By adjusting the structural parameters, the desired phase responses are achieved, as shown in Fig. 3 (a). To maximize the dispersion of scattered EM energy in various directions and enhance the field distribution uniformity within the cavity, GA is employed to optimize the arrangement of the metasurface elements, and the result is illustrated in Fig. 3(b).

*B. Uniformity Evaluation*

To validate the proposed method, the E-field distributions in the microwave heating cavity with and without the metasurface are simulated at 2.45 GHz. Fig. 4 (a) depicts the cavity dimensions and the placement of the metasurface. For a clear comparison, the E-field (|***E***|) at 400,221 mesh points were exported and normalized by their average values. The probability density functions (PDFs) of the normalized E-field, presented in Fig. 4 (b), show that the points with E-field values close to the average are significantly more numerous in the cavity with the proposed metasurface than without it, indicating a substantial improvement in field uniformity. This improvement is illustrated by the total E-field distribution in Fig. 6 (a), where the metasurface eliminates the standing wave pattern observed in the cavity without the metasurface.

Further validation involved heating a piece of potato in the cavity. Supported by a cubic glass with a dielectric constant of 4 [29], the dimensions of the potato and glass are shown in Fig. 4 (a). The temperature distribution in the potato is simulated using COMSOL Multiphysics 6.0 [12], with the relative permittivity of the potato modeled as a function of temperature ($T$): 56.8 - 6.4 × $10^{-3}$ $T^2$ + 0.2 $T - j$ (16.1 – 1 × $10^{-4}$ $T^2$ - 0.108 $T$). Its thermal conductivity coefficient is 0.4 W/m·K, specific heat capacity is 3640 J/kg·K, and density is 1050 kg/m³ [29]. The boundary conditions for EM field and thermal transfer are the same as in [12], [30]. The continuous wave power of 100 W is excited in the waveguide. The initial

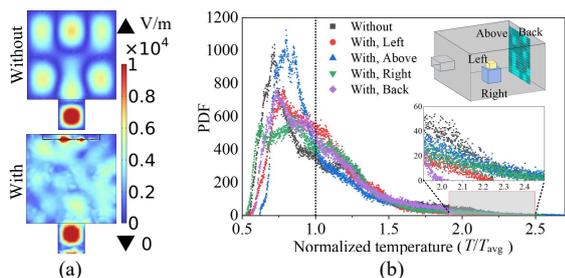

**Fig. 6.** (a) Comparison of the total E-field (|***E***|) distributions at $z$ = 93 mm without and with metasurface in cavity. (b) PDFs of normalized temperature in the potato without and with the metasurface at different walls in the cavity.

temperature is 12 °C, and the total heating duration is 40 seconds.

The temperature distribution inside the potato is normalized by its average value ($T_{ave}$). Fig. 6 (b) presents the PDFs of the normalized temperature at $t$ = 40 s, with and without the metasurface. It shows that the temperature distribution in the potato with the metasurface is more concentrated around the mean value than without it, regardless of the metasurface placement. The heating uniformity is improved greatly.

To quantify this improvement, the coefficient of variation (COV) is employed [8]:

$$COV = \frac{1}{T_{avg} - T_0} \sqrt{\frac{1}{n} \sum_{i=1}^{n} (T_i - T_{avg})^2} \qquad (3)$$

where $T_i$ and $T_{avg}$ denote the discrete temperature within the potato and their average, respectively, $n$ is the number of points (406,702), and $T_0$ is the initial temperature.

The COV without the metasurface is 0.694, which decreases to 0.461, 0.483, 0.425 and 0.524 with the metasurface placed in front of the back, left, right and above walls, respectively (Fig. 6 (b)). The corresponding $T_{avg}$ of the potato with the metasurface is 24.4 °C, 26.7 °C, 20.5 °C, and 28.3 °C, respectively, while $T_{avg}$ is 24.6 °C without the metasurface. The proposed method effectively improves heating uniformity without reducing the overall efficiency, except when the metasurface is placed right. In contrast, the heating efficiency is enhanced when the metasurface is placed in front of the left and above walls, but with a little bit more hotspots (left) or less uniformity (above). In summary, considering COV, average temperature and hotspots, the optimal position for the metasurface is in front of the back wall. These differences in COV and average temperature are mainly caused by the asymmetric design of the metasurface.

## III. EXPERIMENTAL VALIDATION

To experimentally validate the proposed method, a microwave heating system is set up, as shown in Fig. 7. A fiber optic thermometer is used to measure the temperature at the center by inserting the fiber sensor into the potato and record the temperature once per second. An infrared thermal camera is used to immediately record the temperature distribution on the potato's top surface after heating.

Fig. 8 (a) illustrates the simulated and measured surface temperature distributions. Good agreements are observed between them. The temperature distribution on the top surface becomes more uniform with the metasurface. Fig. 8 (b) compares the measured and simulated temporal variations of temperature at the potato center, with and without the metasurface. The experimental data, representing the average temperature from three trials, closely matches the simulation. The discrepancies between the simulation and experimental results stem primarily from differences between the simulation model and the actual heating system.

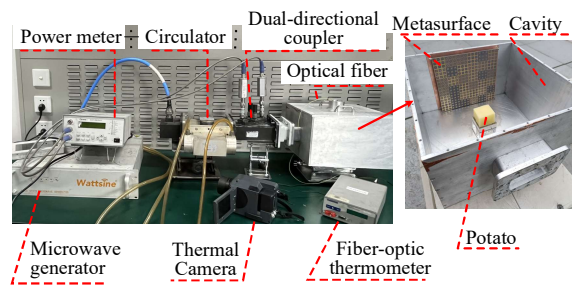

**Fig. 7.** Experimental system for microwave heating.

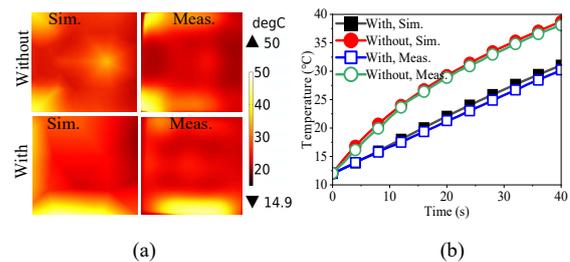

**Fig. 8.** Comparison of potato temperature with and without metasurface in cavity. (a) Temperature distribution on surface, and (b) temperature at the potato center.

Although the temperature rise at the potato center is higher without the metasurface, the average temperature of the entire potato remains nearly the same, as indicated by the simulation results. This suggests that the proposed method effectively mitigates thermal runaway issues.

## VI. CONCLUSION

This study proposes a method to improve heating uniformity in a cavity by introducing a metasurface operating at 2.45 GHz. This approach eliminates the need to redesign the cavity or using multiple feeding sources. With the metasurface being less than 18 mm thick and positioned in front of the wall inside the cavity, the heating space is minimally impacted. The simulation results, including PDFs of the E-field and potato temperature, and COV of temperature, demonstrate significant improvement in heating uniformity, regardless of the metasurface placement. Considering heating efficiency and hotspots, the optimal position can be identified. The simulated potato temperatures closely match the measured ones, validating the proposed method. Moreover, the proposed metasurface can be easily applied to other rectangular microwave heating cavities operating at 2.45 GHz. While the narrow gap in the metasurface structure limits the power tolerance, the proposed method can find applications in a wide range of scenarios.